\documentclass{ws-ijmpd}
\usepackage[super,compress]{cite}
\usepackage{hyperref}
\usepackage{colortbl}
\usepackage{changes}

\begin{document}

\markboth{J.C.N. de Araujo and H.G.M. Fortes}
{Solving Tolman-Oppenheimer-Volkoff equations in $f(T)$ gravity}

%%%%%%%%%%%%%%%%%%%%% Publisher's Area please ignore %%%%%%%%%%%%%%%
%
\catchline{}{}{}{}{}
%
%%%%%%%%%%%%%%%%%%%%%%%%%%%%%%%%%%%%%%%%%%%%%%%%%%%%%%%%%%%%%%%%%%%%

\title{Solving Tolman-Oppenheimer-Volkoff equations in $f(T)$ gravity: a novel approach applied to some realistic equations of state}

\author{J. C. N. Araujo$^{\star}$ and H. G. M. Fortes$^{\dagger}$}

\address{Divis\~{a}o de Astrof\'{i}sica, Instituto Nacional de Pesquisas Espaciais, \\
Avenida dos Astronautas 1758 \\ S.J. Campos, SP 12227-010, Brazil \\ $^{\star}$jcarlos.dearaujo@inpe.br \\ $^{\dagger}$hemily.gomes@inpe.br}

\maketitle

\begin{history}
\received{Day Month Year}
\revised{Day Month Year}
\end{history}

\begin{abstract}
There are many ways to probe alternative theories of gravity, namely, via: experimental tests at solar system scale, cosmological data and models, gravitational waves and compact objects. In the present paper we consider a model of gravity with torsion $f(T)$ applied to compact objects such as neutron stars (NSs) for a couple of realistic equations of state (EOS). To do so we follow our previous articles, in which we show how to model compact stars in  $f(T)$ gravity by obtaining its corresponding Tolman-Oppenheimer-Volkof equations. In these modeling of NS in $f(T)$ gravity presented here, we calculate, among other things, the maximum mass allowed for a given realistic EOS, which would also allow us to evaluate which models are in accordance with observations. The results already known to General Relativity must be reproduced to some extent and, eventually, we can find models that allow higher maximum masses for NSs than Relativity itself, which could explain, for example, the secondary component of the event GW190814, if this star is a massive NS.
\end{abstract}


\ccode{PACS numbers: 04.50.-h; 04.50.Kd; 97.60.Jd}

%PACS ver, e.g., https://ufn.ru/en/pacs/
% 04.50.-h  alternative theories of gravity
% 04.50.Kd	Modified theories of gravity
% 97.60.Jd Neutron stars

%\tableofcontents

\section{Introduction}
\label{int}

The nuclear matter inside a star is characterised by an equation of state (EOS) whose form is a result of several contributing factors. For this reason, there are different candidates for the most accurate EOS in order to describe compact objects which can be derived from the phenomenological nuclear interactions and, more recently, using elements of the chiral effective field theory (EFT) based on quantum chromodynamics (QCD) (see, e.g., Ref. \refcite{Greif}). 

The EOS of dense matter is fundamental in modeling neutron stars (NSs), since it determines, for example, the radius, the mass, the moment of inertia and the tidal deformability. Constraints on these parameters can be obtained from theoretical and experimental nuclear physics, and from different (multi-menssenger) astrophysical observations. It is worth stressing that the gravitational waves (GWs), since the GW170817 event, are now part of multi-messenger astrophysics. We refer the reader to Ref. \refcite{latimer} for a review concerning all these issues.

Recall that, GWs emitted during the merger of NS-NS binaries also depend on the EOS, since the gravitational waveform depends on the gravitational tidal deformation that one star produces in the other \cite{GW170817}. Consequently, data from GW observations that come from NS-NS mergers can provide information about high density nuclear EOSs, i.e., larger than the saturation density ($\rho_{nuc}=2.8\times 10^{14} g \ cm^{-3}$). On the other hand, laboratory nuclear experiments, for example, can only constrain EOSs at densities  around or below $\rho_{nuc}$. GW170818 can also be used to constrain the neutron star radius\cite{Bauswein}.

{It is worth mentioning that NICER (acronym for the Neutron Star Interior Composition Explorer) data can be used to constrain the EOS of neutron stars and provide estimates for their masses and radius, see, e.g., Refs. \refcite{Raaijmakers} and \refcite{Alti}.}

Any kind of additional information and constraints about the EOS are useful in order to make possible to rule out some of them. The fact, for instance, of existing NSs with greater mass than we expected from the theoretical point of view is a fundamental eligibility criteria for the EOS. Additionally, complementary constraints have been provided in the literature from nuclear physics, moment of inertia measurements and observational data in general \cite{Rav,Lyne,Lat,Heb,Greif,Tsang}.

The EOSs that are considered adequate to the constraints from theory and observations, together with the Tolman-Oppenheimer-Volkoff (TOV) equations \cite{TOV}, can be used to model the structure of a spherically symmetrical object that is in hydrostatic equilibrium. The TOV equations are derived from the theory of gravity considered. Among the available alternative theories of gravity, the models with torsion have stood out. The so-called $f(T)$ models have a function of torsion $T$ as the Lagrangian density, instead of the curvature $R$. These models have, in general, simpler field equations than those obtained in $f(R)$ and have presented interesting cosmological and astrophysical solutions \cite{review2,Karami}.

In Ref. \refcite{FA2}, it was used the polytropic equation of state to describe the structure of a spherically symmetrical object in gravitational equilibrium in a model with torsion, namely, $f(T,\xi) = T + \xi T^2$, where $\xi$ is a free parameter. In the present work, we will perform the numerical calculations in a analogous way what was done in Ref. \refcite{FA2} for this family of alternative models using, this time, realistic EOSs. 

The TOV equations correspondent to $f(T,\xi)$ were obtained in Ref. \refcite{FA1} using a novel approach without any restriction on the metric functions or perturbative calculations. Therefore, Ref. \refcite{FA1} is the basis for the present paper.

Modeling NSs with realistic equations of state in any gravity theory is essential, since only so it is possible to constrain both. Notice that if the secondary component of the event GW190814 is a massive NS, GR can hardly explain such a star, even considering rotation. Therefore, it is essential to investigate what alternative gravity has to say in this regard.

{There are several relevant studies in the literature of a series of alternative theories related to modeling neutron stars. For a recent review, see \refcite{olmo}. In addition, there are a number of studies on neutron star modeling in $f(R)$ worth paying attention\cite{Yaza,Feola,Capo2016,Asta}.
Morevover, we can also mention the neutron star modeling on scalar-tensor gravity\cite{Odintsov21,Odintsov22}.}

In Section \ref{sec 2}, we present the main results obtained in Ref. \refcite{FA1} concerning the model $f(T,\xi)$. In Section \ref{sec 3}, we proceed analogously to Ref. \refcite{FA2} in the numerical calculations in order to obtain the ``Mass × Radius'' and ``Mass × Central Density'' curves, from which one can obtain, for example, the maximum mass allowed for some realistic and representative EOSs. Finally, the main conclusions are presented in Section \ref{sec 4}. 

\section{Basic Equations}\label{sec 2}

In our previous papers \cite{FA2,FA1}, we show in detail how to obtain the set of differential equations necessary to model compact stars in a particular $f(T)$ gravity adopting polytropic EOSs. In the present paper, we consider the same set of equations, but now solved for realistic EOSs.

The equations of motion derived from the action for the $f(T)$ theory are as follows \cite{FA1}:
\begin{eqnarray} %\hspace{-1 cm}
  &\,& 4\pi P = -\frac{f}{4} +\frac{f_T\,e^{-B}}{4r^2} \left( 2-2\,e^B+r^2e^B T-2r\,A' \right) 
\label{026}\\ %\hspace{-1 cm}
&\,&4\pi\rho = \frac{f}{4} -\frac{f_T\,e^{-B}}{4r^2}\left( 2-2\,e^B+r^2e^B T-2r\,B' \right) %\nonumber \\ 
- \frac{f_{TT}\,T'e^{-B}}{r} \left(1-e^{B/2}\right) 
\label{025}\\ %\hspace{-1 cm}
    %\hspace{-1 cm}  %+\nonumber \\
  &\,&f_T \Big[ 4\,e^A-4e^Ae^B-e^Ar^2A'^2+ 2\,e^Ar\,B' +e^Ar\,A' \left( 2+r\,B' \right) -2r^2e^A A'' \Big] +\nonumber \\
&\,& +  f_{TT}\Bigl[ -4\,e^A r\, T'(1-e^{B/2} \sin{\gamma})  -2r\,A' e^Ar\,T'\Bigr]=0
\label{027}
\end{eqnarray}
where $f_T=\frac{\partial f}{\partial T}$ and $f_{TT}=\frac{\partial f_T}{\partial T}$ and we have considered $A$, $B$ time independent.

{Recall that we are considering spherical stars, since rotation is not taken into account. Thus, we adopt the spherically symmetric metric} {$ds^2=e^{A(r)}\, dt^2-e^{B(r)}\, dr^2-r^2\, d\theta^2-r^2 \sin ^2 \theta \, d\phi^2$.}

The form chosen for $f(T)$ in Ref. \refcite{FA1} was 
\begin{eqnarray}
 f(T)=T+\xi \, T^2 \ ,\label{fT}
\end{eqnarray}
where $\xi$ is an arbitrary real. Notice that, for $\xi=0$, the results from the Teleparallel equivalent  of  General  Relativity can be retrieved.

{This specific choice has different motivations. Among them, it is the simplest one and it is inspired in the Starobinsky model in $f(R)$ gravity, which has this same functional form. However, particularly following Ref. \refcite{FA1}, it is straightforward to consider other functional forms for $f(T)$. And the corresponding numerical scheme to solve the set of equations is basically the same presented later on this paper.}

From the Eqs. (\ref{026})-(\ref{027}), it was obtained  $P$ and $B'$, namely,
\begin{eqnarray}
P &=& \frac{{c^4}\, {\rm e}^{-B}}{8\pi {G} r^4}\Big\{{r}^{2} \left(1-{{\rm e}^{B}} \right) +2{{\rm e}^{B}}({{\rm e}^{-1/2\,B}}-1)^3(3{{\rm e}^{-1/2\,B}}+1)\xi+
\nonumber \\
&+& r\Big[r^2 +12({{\rm e}^{-1/2\,B}}-1)^2\xi\Big] A'+ 2r^2 \xi({{\rm e}^{-1/2\,B}}-1)(3{{\rm e}^{-1/2\,B}}-1) \, {A'}^2\Big\}%\hspace{0.5cm}
\label{press}
\end{eqnarray}

\par and

{
\begin{eqnarray} %\hspace{-2 cm}
B'&=& -\frac{{{\rm e}^{B}}}{r}\bigg\{{r}^{4}\big(1-{{\rm e}^{-B}}-8\,\pi \,{r}^{2}\rho\big)
%+\nonumber \\
-96\pi\,{r}^{4}\rho ({{\rm e}^{-1/2\,B}}-1)^2\xi+ 
\nonumber \\    %\hspace{-2 cm}
&-& 6\,{r}^{2}({{\rm e}^{-1/2\,B}}-1)^3(5+3{{\rm e}^{-1/2B}})\xi
%+\nonumber \\
-8({{\rm e}^{-1/2\,B}}-1)^5(11+9{{\rm e}^{-1/2B}})\xi^2 +
\nonumber \\    %\hspace{-2 cm}
&-& 8\,{r}^{3}{{\rm e}^{-1/2\,B}}({{\rm e}^{-1/2\,B}}-1)\Bigl[8\,\pi \,{r}^{2}\rho
%+ \nonumber \\
+({{\rm e}^{-1/2\,B}}-1)(2{{\rm e}^{-1/2\,B}}+1)\Bigr]\,A'\xi+
\nonumber \\    %\hspace{-2 cm}
&-& 16\, r\, {{\rm e}^{-1/2\,B}}({{\rm e}^{-1/2\,B}}-1)^4(9{{\rm e}^{-1/2\,B}}+5)\,A'\xi^2
%+\nonumber \\
+2\,r^4 {{\rm e}^{-B}}({{\rm e}^{-1/2\,B}}-1)^2  \,{A'}^2\xi+
\nonumber \\    %\hspace{-2 cm}
 &-& 8\,r^2 {{\rm e}^{-B}}({{\rm e}^{-1/2\,B}}-1)^3 (9{{\rm e}^{-1/2\,B}}-1)\,{A'}^2\, \xi^2\bigg\}
\bigg/ % denominador - CORRIGIDO
%\nonumber \\ 
 \bigg\{ r^4+16\,{r}^{2}({{\rm e}^{-1/2\,B}}-1)^2\xi+
\nonumber \\   %\hspace{-2 cm}
&+& 48\,({{\rm e}^{-1/2\,B}}-1)^4\xi^2
%+\nonumber \\ 
+16\,{r}^{3}{{\rm e}^{-1/2\,B}}({{\rm e}^{-1/2\,B}}-1) A'\xi +
\nonumber \\   %\hspace{-2 cm}
&+& 96\,r\,{{\rm e}^{-1/2\,B}}({{\rm e}^{-1/2\,B}}-1)^3 A'\xi^2 
%+\nonumber \\
   +48\,{r}^{2}{{\rm e}^{-B}}({{\rm e}^{-1/2\,B}}-1)^2
 {A'}^2\xi^2 \bigg\} %\hspace{0.56 cm}
\label{BL}
\end{eqnarray}
}

{A differential equation for $A''$ was also found. However, it was not necessary to use that in the numerical integration. Instead, we have used the conservation equation,
\begin{equation}
    2 P' + (P+\rho)A' = 0  \ ,
\label{ce1}    
\end{equation} 
which also holds in $f(T)$ gravity (see Ref. \refcite{Bohmer} for details) and we have realised that use it rather than $A''$ would be more productive, since this makes the numerical solution much easier to deal with.
We refer the reader to Refs. \refcite{FA2} and \refcite{FA1}, where further details are provided. It is worth mentioning that Ref. \refcite{Ilijic18} did not realised that the appropriate combination of their equations leads to the {\it conservation equation}.}

In addition, the mass equation is given by: 
\begin{eqnarray}
\frac{dm}{dr}=4\pi \rho  r^2\ .
\label{dmdr}
\end{eqnarray}

{Notice that the above equation is just like the one of TOV GR. The mass of compact objects in General Relativity, obtained through the Tolman-Oppenheimer-Volkov equations, is a well-defined quantity, whereas, in alternative gravity, there may be ambiguity in its definition\cite{olmo}. As an alternative way of measuring the mass of a star, for example, it is possible to consider the total number of particles $N$ by integrating the density of the number of particles $n=dN/dV$ over the interior of the star, which is closely related to the total rest mass. Despite all this, the form in (\ref{dmdr}) is also adopted {implicitly or explicitly} by other authors (see, e.g., Refs. \refcite{Ilijic20} and \refcite{lcz}) for the calculation of mass in $f(T)$ gravity. On the other hand, this topic is considered by some authors an open problem (see, e.g., Refs. \refcite{olmo} and \refcite{Ilijic18}). In Refs. \refcite{FA2} and \refcite{FA1} we also briefly discuss such an issue.} 

Also, note that the energy density $\rho$ in the equations above is related to the pressure $P$ through the EOS. 

Obviously, $\xi=0$ gives $f(T)=T$, which is nothing but GR  (see Ref. \refcite{FA1} for detail). Thus, the system of Eqs. (\ref{press}), (\ref{BL}), (\ref{ce1}) and (\ref{dmdr}) for $\xi=0$ is equivalent to TOV GR. {Notice that $A(r)$ does not appear in this system of equations, therefore one does not need to obtain it. In fact, what appears in this system is $A'$, its first derivative, which is algebraically related to $P(r)$ and $B(r)$ via Eq. \ref{press}. {Thus, one does not need a differential equation for either $A'$ or $A''$, since $A'$ can be obtained via Eq. \ref{press}.}}

Although it is not an observable, the total rest mass $M_0$ is an interesting quantity to calculate, since it can also be used to compare different theories of gravity. For example, for given EOS and central density, different theories provide different values of $M_0$. Recall that the total rest mass $M_0$ is obtained from
\begin{eqnarray}
\frac{dm_0}{dr}=4\pi \rho_0 \,e^{B/2} r^2
\label{dm0dr}
\end{eqnarray}
where $\rho_0$ is the rest mass density.

{It is worth mentioning that we have followed here the usual non-covariant formulation. However, the covariant formulation, which does not depend on the choice of any particular tetrad, was adressed in Ref. \refcite{Krssak} where the authors showed that this formulation and the non-covariant one, with the tetrad adopted here, yield the same set of differential equations.}

\section{Numerical Calculations and Discussions}\label{sec 3}

In this section we present the main results of our numerical calculations. To do so, we have integrated the set of equations of the previous section, namely, (\ref{press}), (\ref{BL}), (\ref{ce1}) and (\ref{dmdr}), 
to model the NSs for given EOSs.
 
In addition, to model a NS, we choose a central density, $\rho_c = \rho (r=0)$, and adopt an EOS to obtain the central pressure $P_c = P(r=0)$. Moreover, the central boundary conditions,
\begin{equation}
    m = 0 \quad {\rm and} \quad B = B_c \quad {\rm at} \quad r =0,
\end{equation}
are set to integrate the system of differential equations in order to obtain $m(r)$, $P(r)$, $\rho(r)$ and $B(r)$. The radius $R$ of the NS is the value of $r$ for which $P(r) = 0$. That is, starting the integration at $r=0$ and following it up to a certain $r$ where $P(r)=0$. The mass of the NS is then given by $ M = m(R)$. 

Notice that, in comparison to GR, there is an additional equation, namely, the equation (\ref{BL}). Based on GR one concludes that $B_c ~=~0$ (see, e.g., Ref. \refcite{FA2}). {Also, in  $f(R)$, for example, this central boundary condition is  the same.}

{Still regarding to $f(R)$, another boundary condition is necessary, namely, the central value of the Ricci scalar $R$ ($R_c$), since $R$ is obtained via a differential equation. Then, $R_c$ is chosen such that $R$ goes to zero at infinity, i.e.,  the asymptotic flatness must be satisfied.}

{Here it is not necessary to provide a central value of the torsion $T$, as there is no differential equation for it. The torsion is given by}
{\begin{equation}
T(r) = \frac{2\, e^{-B}}{r^2} 
\Bigl[ 1 + e^B - 2\,e^{B/2} + 
 r\, A' \left(1-e^{B/2}\right)
\Bigr]
  \label{008}
\end{equation}}
(see, e.g., Ref. \refcite{FA1}). Our calculations show that $B$ and $A'$ go to zero at r $\rightarrow \infty$ which implies that $T$ goes to zero. Consequently, it is not necessary to impose any asymptotic flatness condition, as it is naturally satisfied.

{The energy density profile is affected by the metric functions $A$ and $B$, which in turn are affected by the particular $f(T)$ gravity considered. This can be seen from equation (\ref{BL}) which holds inside and outside the star, since $\rho$ goes to zero smoothly. Then, the behaviour of $B$ outside the star, for example, is affected by the particular $f(T)$ considered. We do not impose any functional form for $B(r)$, as some authors do.\cite{Oiko21,Oiko22} This function is always obtained via the numerical integration of equation (\ref{BL}) in our calculations. Consequently, the Schwarzschild metric is not assumed as a exterior solution in our approach.}

Before proceeding, it is worth mentioning that in order to integrate the system of differential equations to model the NSs, we wrote a code in Python for this purpose and used a Runge-Kutta integrator. In all calculations, the EOSs used in our Python code are in table form. To manipulate the table of EOSs we use a cubic spline routine of the Python library itself.

As already mentioned, it is  well known that the EOS of NSs is an open question. That is why there is in the literature a host of NS EOSs proposals (see, e.g., Refs. \refcite{Greif}, \refcite{Heb} and \refcite{Ozel}). Here, we adopt a couple of them without entering in the corresponding nuclear theory. {The basic idea is to see how different EOSs behave under the particular $f(T)$ gravity considered here, especially with regard to the maximum masses.} {We do not intend to constrain the EOSs in the present paper. A study addressing this issue is left for a future publication, where not only the EOSs can be constrained but also the $\xi$ parameter of the $f(T)$ gravity studied here.}

In Ref. \refcite{Heb} (see also Ref. \refcite{Greif}), the authors propose three representative EOSs, namely, soft, intermediate and stiff, with the aim to characterise the different classes of EOSs.  This is an interesting idea since it gives a general idea on how radii and masses behave depending on the stiffness of the EOS. Here, in particular, we adopt their soft and stiff EOSs (Tables 5 - 7 of Ref. \refcite{Heb}).
Their soft EOS has a typical radii of $\sim 10$ km and maximum mass of $\sim 2$ M$_\odot$, whereas their stiff EOS has typical radii and maximum mass of $\sim 14$ km and $\sim 3$ M$_\odot$, respectively. 
{In general, it is possible to say that the harder the EOSs are, the greater the radii and the higher the masses are.}

An interesting way to compare alternative theories to GR is via sequences such as ``$Mass$ $\times$ $R$'' or ``$Mass$ $\times$ $\rho_c$'', since it is possible to see how radii and maximum masses are in a theory other than GR. Moreover, for a given $\rho_c$, alternatives theories could predict different masses and radii of those predicted by GR and this can be easily seen in the sequences of models mentioned. 

\begin{figure*}
\begin{center}
\includegraphics[scale=0.25]{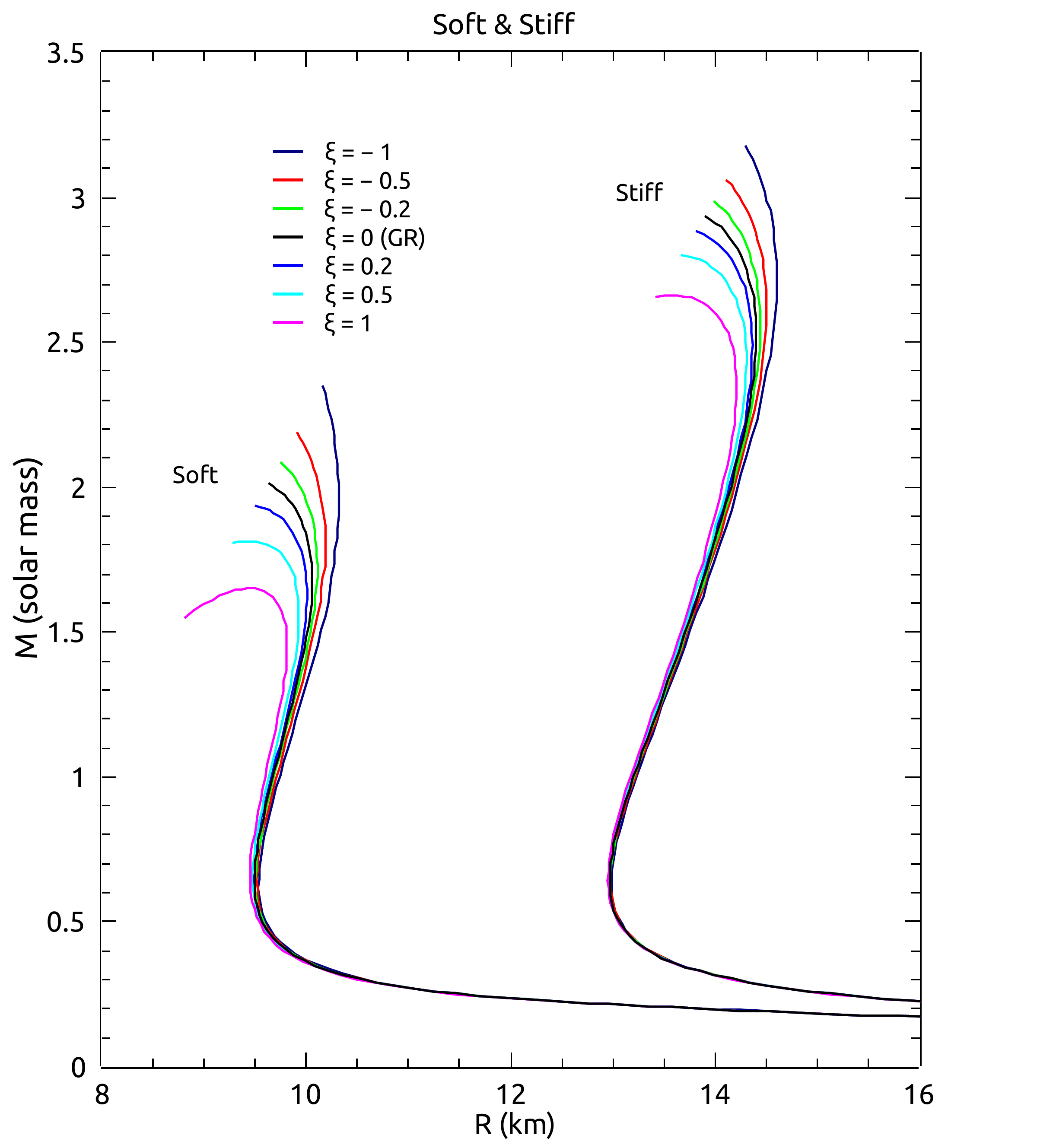}
\includegraphics[scale=0.25]{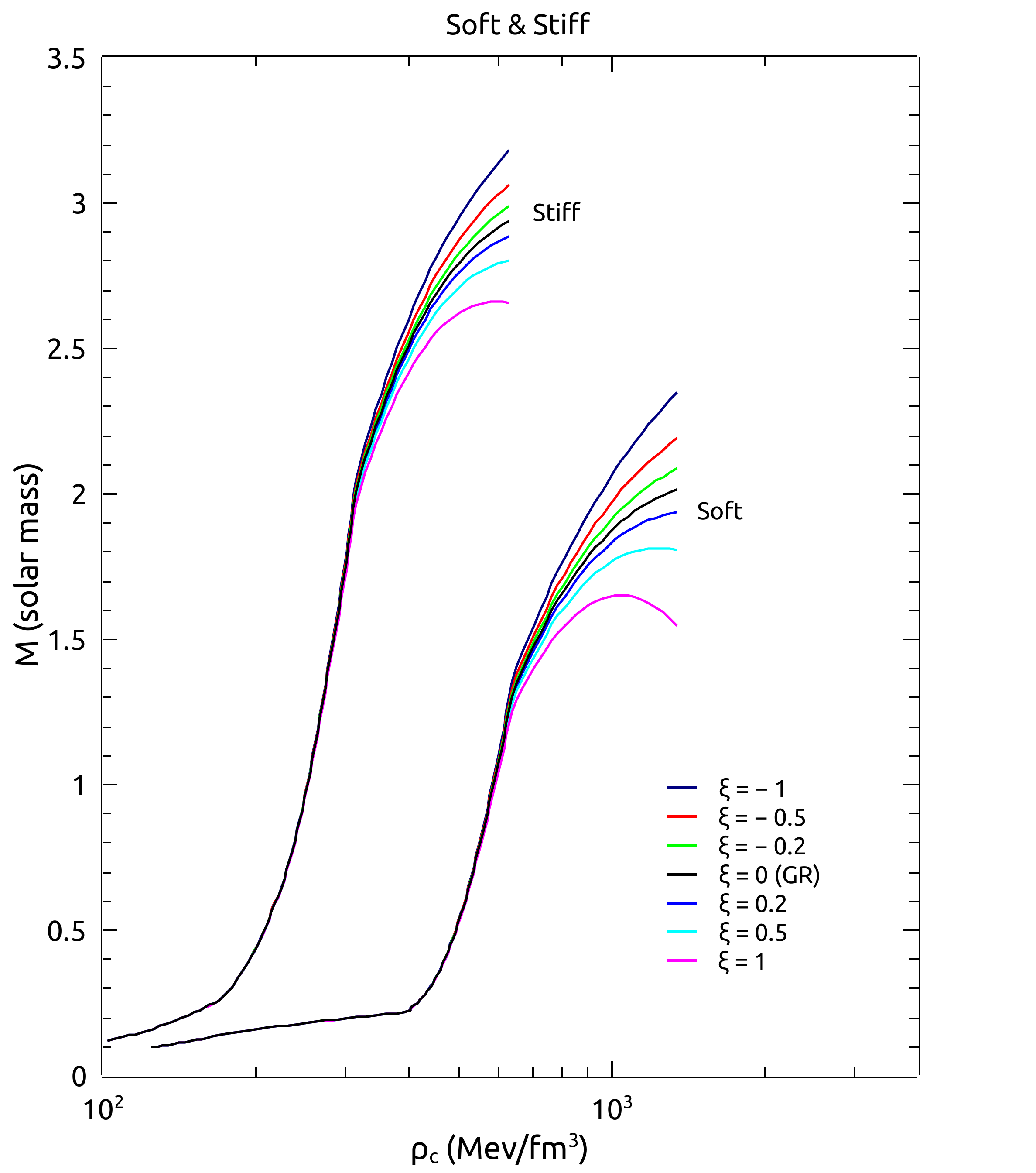}
    \caption{Left (right): sequences of mass $M$ as a function of the radius $R$ (the central energy density $\rho_c$)  for the soft and stiff EOSs\cite{Heb}  and different values of $\xi$  given in units of the square of the gravitational radius of the Sun ($r_{gs} = 2GM_\odot/c^2 \simeq 2.95\, {\rm km})$.
    }
    \label{SSMR}
\end{center}
\end{figure*}

\begin{figure*}
\begin{center}
    \includegraphics[scale=0.25]{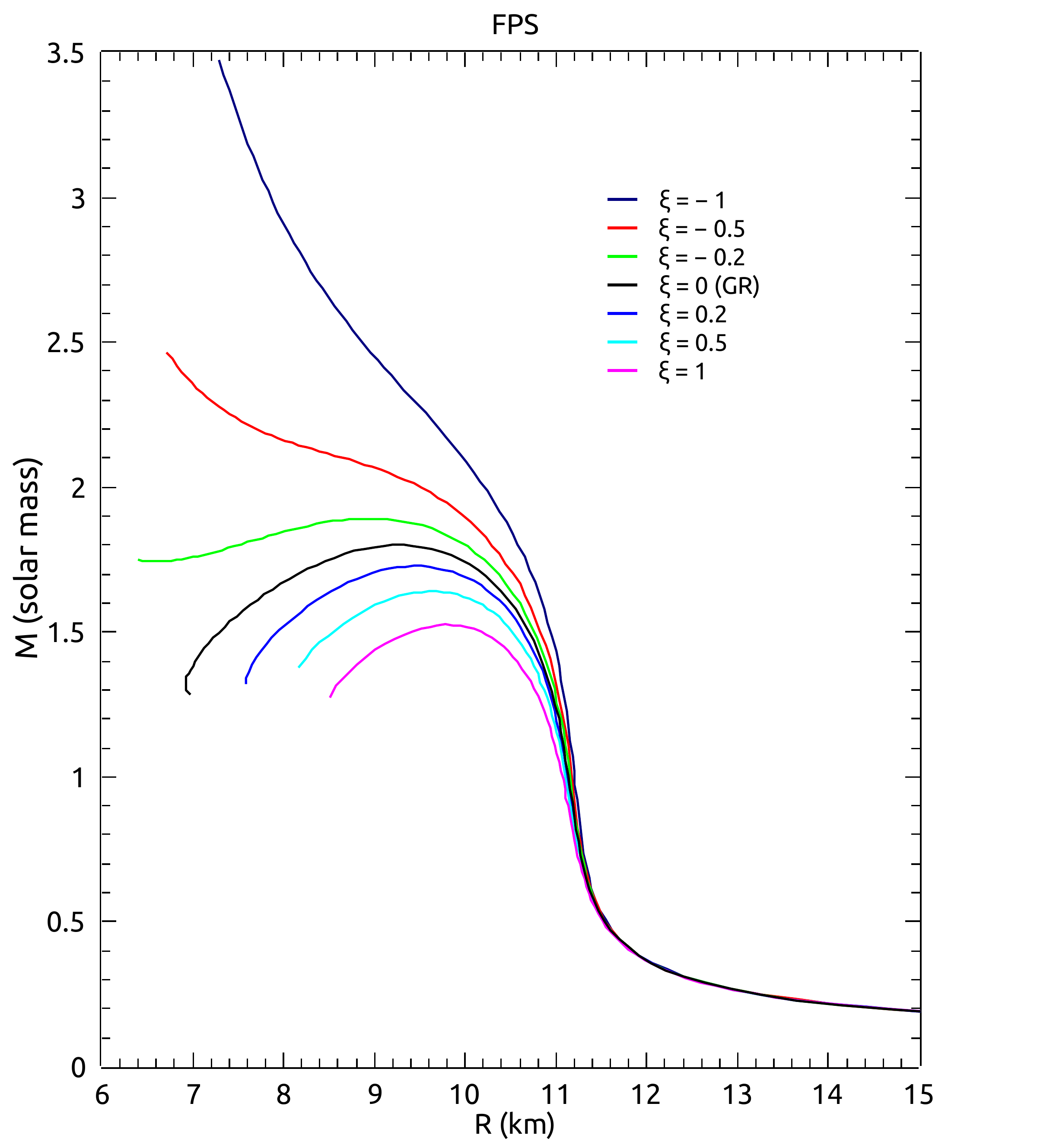}
    \includegraphics[scale=0.25]{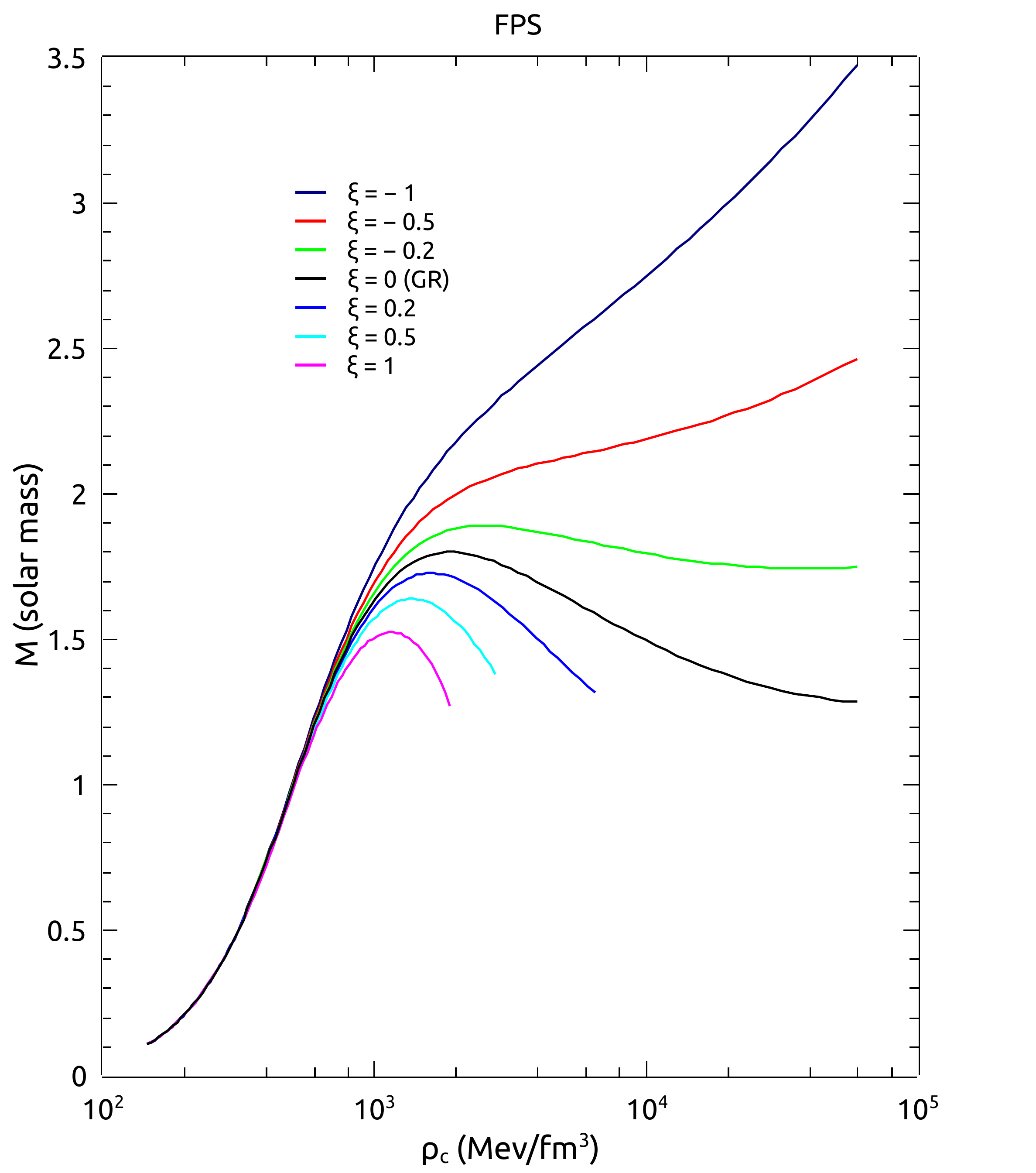}
    \caption{The same as Fig. \ref{SSMR} now for FPS.}
    \label{FPSMR}
\end{center}
\end{figure*}

\begin{figure*}
\begin{center}
    \includegraphics[scale=0.25]{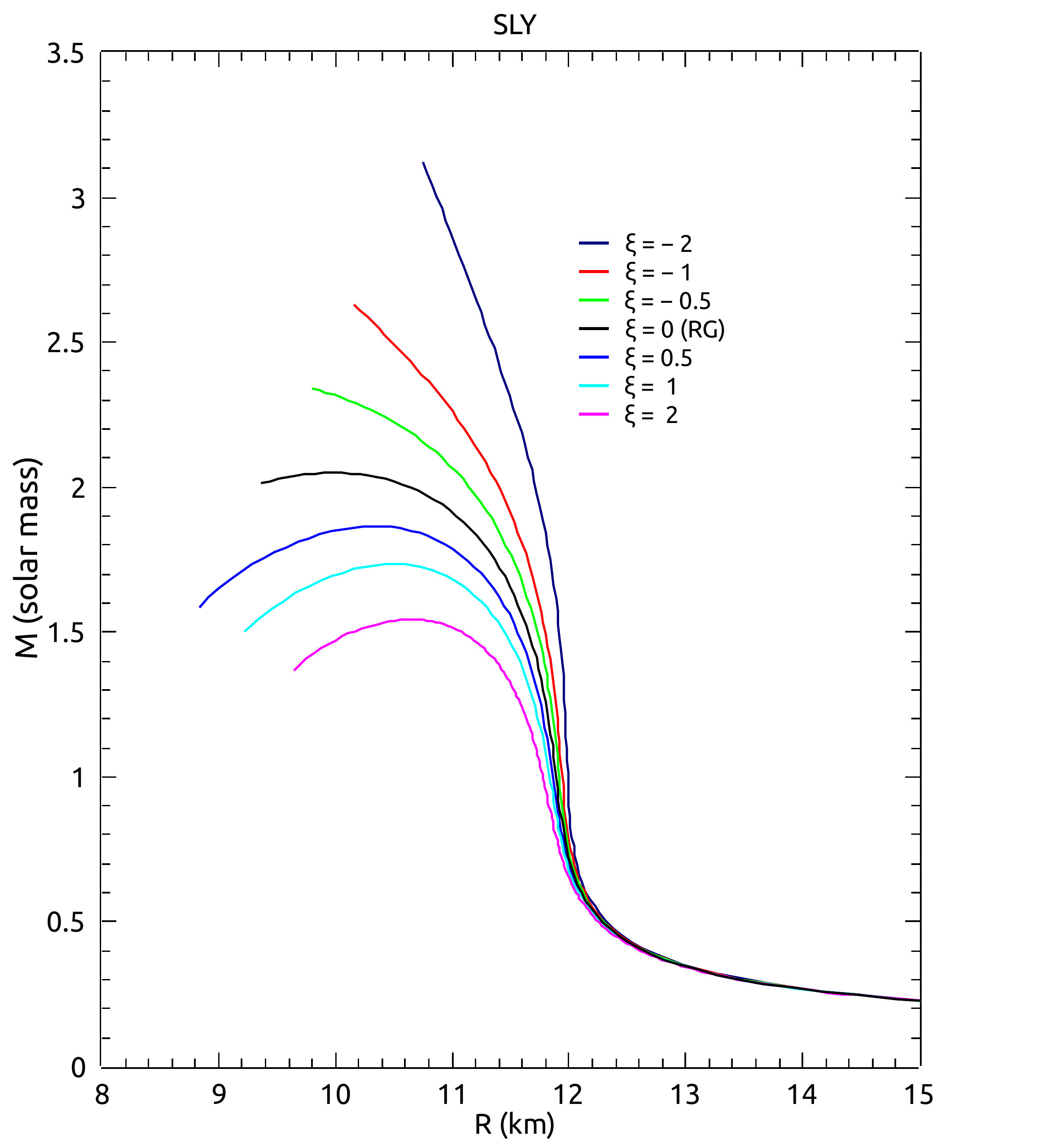}
    \includegraphics[scale=0.25]{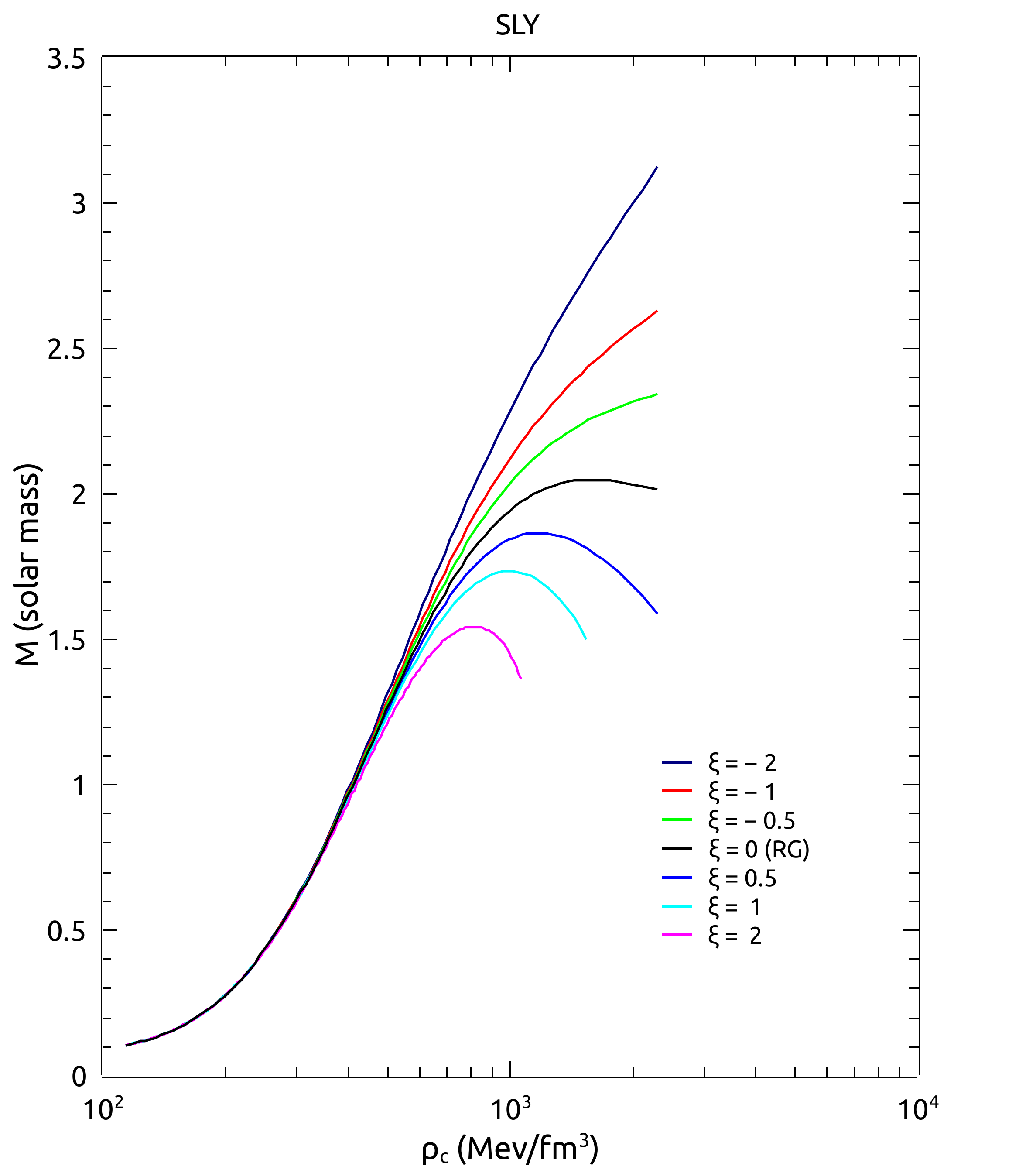}
    \caption{The same as Fig. \ref{SSMR} now for SLY.}
    \label{SLYMR}
\end{center}
    
\end{figure*}

\begin{figure*}
\begin{center}
    \includegraphics[scale=0.25]{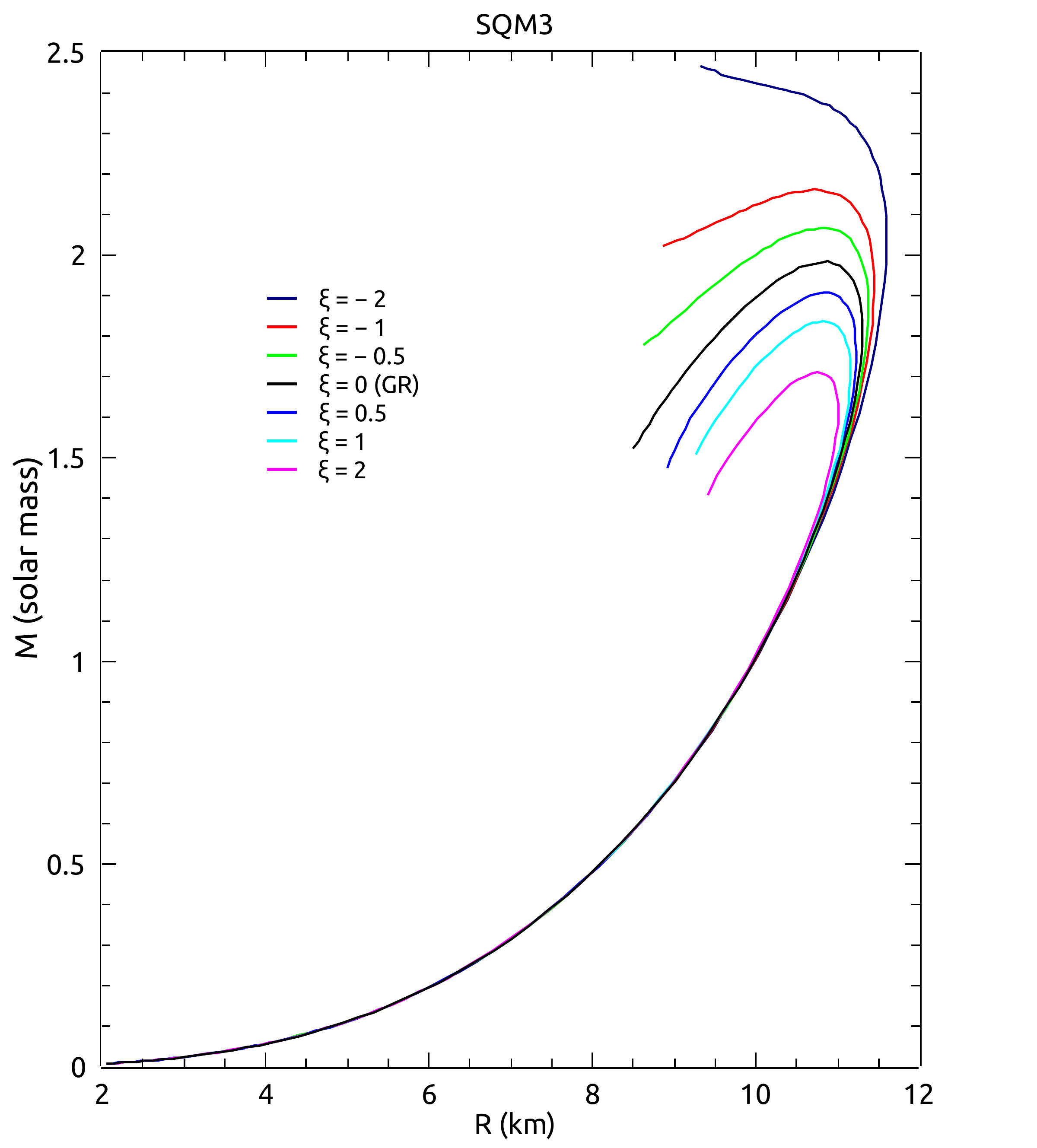}
     \includegraphics[scale=0.25]{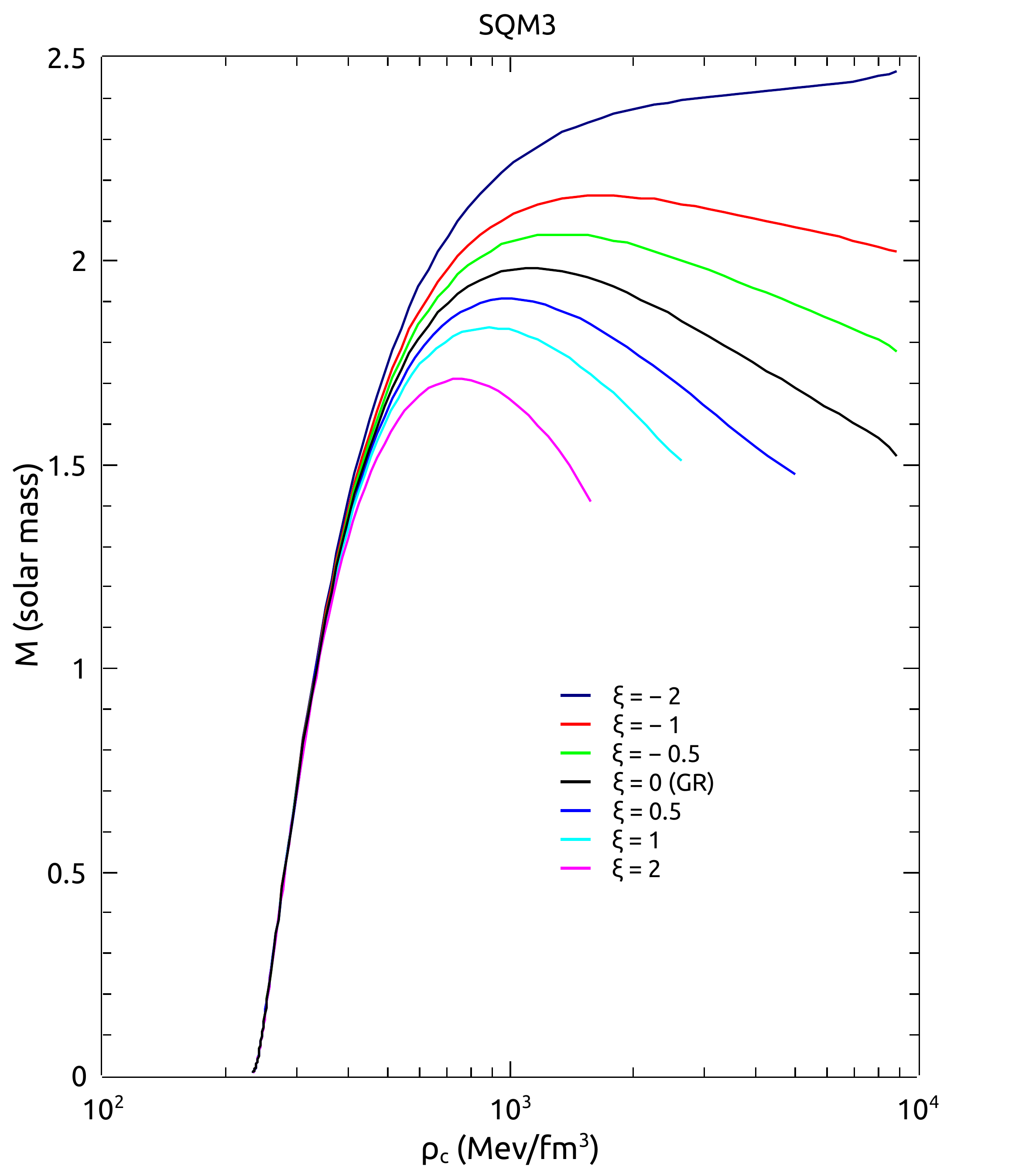}
    \caption{The same as Fig. \ref{SSMR} now for SQM3.} \label{SQM3MR}
\end{center}
\end{figure*}

To begin with, Fig. \ref{SSMR}, on the left panel, shows sequences ``$Mass$ $\times$ $R$'' for the representative soft and stiff EOSs presented in Ref. \refcite{Heb} for different values of $\xi$, which is given in units of the square of the gravitational radius of the Sun ($r_{gs} = 2GM_\odot/c^2 \simeq 2.95\, {\rm km})$.

An obvious fact is that for $\xi > 0$ ($\xi < 0$) the sequences have maximum masses smaller (greater) than that of GR. Also, the greater (smaller) $\xi$ is, the smaller (greater) the maximum mass. Our results suggest that the maximum mass decreases (increases) monotonically for increasing (decreasing) $\xi$. 
{As a general conclusion, maximum masses much greater than those predicted by GR can be obtained for $\xi < 0$.}
The radii are not significantly affected by different values of $\xi$. The compactness, however, is sensitive to $\xi$. The smaller $\xi$ is, the greater the compactness is, and vice-versa. 

The sequences ``Mass $\times$ $\rho_c$'' also provide interesting information; see
the right panel of Fig. \ref{SSMR}.
{For $|\xi | \leq 1 \, r_{gs}^2$, the differences to GR are evident only for $\rho_c > 600 \; (300)\;{\rm Mev/fm^3}$ for the soft (stiff) EOS.} For a given $ \rho_c $ above these values, note that the larger (smaller) $\xi $, the smaller (larger) the mass. {Depending on the value of $\xi$, the mass can easily exceed 3 M$_\odot$. We will discuss this issue a little more later on.}

A well studied soft EOS worth considering is the so-called FPS \cite{FPS}. For this EOS, we also take $|\xi | \leq 1 \, r_{gs}^2$.  
In Fig. \ref{FPSMR}, on the left panel, the ``Mass $\times$ R'' sequences
for FPS EOS for different values of $\xi$ are shown. As compared to the representative soft and stiff EOSs, the FPS is  more sensitive to $\xi$. For GR, the maximum mass is $\simeq 1.8 {\rm M_\odot}$, whereas for $\xi  = - 1 \, r_{gs}^2$ we obtain {a maximum mass much higher, namely,} $\simeq 3.5 {\rm M_\odot}$. On the other hand, for $\xi  = 1 \, r_{gs}^2$ the maximum is $\simeq 1.5 {\rm M_\odot}$. The ``Mass $\times$ $\rho_c$'' sequences are shown on the right panel of Fig. \ref{FPSMR}. For the values of $\xi$ considered, the differences to GR appear only for $\rho_c > 600 \;{\rm Mev/fm^3}$. 

Another EOS worth considering is SLY \cite{SLY}. This EOS is less sensitive to $\xi$ than FPS. Then, we take $|\xi | \leq 2 \, {\rm r_{gs}^2}$. In Fig. \ref{SLYMR}, the ``Mass $\times$ R'' and ``Mass $\times$ $\rho_c$'' sequences are shown. 
Notice that for $\xi =  -2\, (2) \, {\rm r_{gs}^2}$, the maximum mass is of $\simeq 3.1 \,(1.5)\, {\rm M_\odot}$. Recall that for GR the maximum mass is of $\simeq 2.0 {\rm M_\odot}$. Notice that for $|\xi | \leq 2 \,{\rm r_{gs}^2}$, the differences to GR appear only for $\rho_c > 500 \;{\rm Mev/fm^3}$. 

Finally, we consider the SQM3 EOS\footnote{Tables with SQM3, SLY and FPS EOSs can be found, e.g., at \url{http://xtreme.as.arizona.edu/NeutronStars/}.}, which is made of quarkionic matter. This EOS is less sensitive to $\xi$ than the FPS EOS. Whereas for GR the maximum mass is of $\simeq 2.0 {\rm M_\odot}$, for $\xi =  -2\, (2) \, {\rm r_{gs}^2}$ the maximum mass is of $\simeq 2.5 \,(1.7)\, {\rm M_\odot}$. For $|\xi | \leq 2 \;{\rm r_{gs}^2}$, the differences to GR appear for $\rho_c > 400 \;{\rm Mev/fm^3}$; see Fig. \ref{SQM3MR}.

{We can conclude that although the EOSs concerned here show different behaviours regarding to the $f(T)$ considered in the present paper, it is clear that maximum masses much greater than those predicted by GR can be obtained.}

Notice that, depending on the value of $\xi$, the maximum mass of a NS can be $>$ 2.5 M$_\odot$, even without considering rotation. In Ref. \refcite{Breu} it is studied how much the maximum mass is modified by rotation. They found, considering GR, that the maximum mass can be increased in up to $\sim 20\%$. If in the particular $f(T)$ studied here the maximum mass is affected in a similar amount, it is possible to say that maximum masses $>$ 3 M$_\odot$ can be obtained even for soft EOSs.

{As seen, for different EOSs, and for the particular $f(T)$ studied, masses of 3M$_\odot$ can be exceeded by many EOSs, even if the associated speed of sound does not exceed the speed of light. In fact, the more negative $\xi$ is, the greater the maximum mass. However, it is important to study the stability of these stars,
since a limit mass for $f(T)$ must exist.
Recall that in scalar tensor gravity\cite{Sotani}, as well as in $f(R)$ gravity\cite{AstaPLB,Asta136}, there is a maximum mass. This study, however, is left for a future work because it is outside the scope of the present one. }

In Ref. \refcite{gw190814}, it is reported the event GW190814, which is a result of a coalescence involving a 22.2–24.3 M$_\odot$ black hole (BH) and a compact object with a mass of 2.50–2.67 M$_\odot$, which can be either a very massive NS or the lightest black hole discovered. We find that, the $f(T)$ here studied could well explain such a putative high mass NS, depending obviously on the value of $\xi$,  even considering a soft EOS and no rotation. Notice that the GW170817 event favoured soft EOSs (see, e.g., Ref. \refcite{GW170817}).

{Constraints on the radii and equation of state of neutron stars can be obtained by analyzing gravitational wave data from events such as GW170817\cite{GW170817,Bauswein}. In addition, NICER data can also be used for this purpose\cite{Raaijmakers}.}

{Generally speaking, GW170817 favours soft EOSs\cite{GW170817,Bauswein}, such as, for example, the SLY considered here, which is consistent with the constraints obtained in a study involving NICER's data of PSR J0740+6620\cite{Raaijmakers}.}

{Positive (negative) values of $\xi$ {\it work like a softener (stiffener)} of EOSs  (see, e.g., the left panels of figures \ref{SSMR}, \ref{FPSMR}, \ref{SLYMR} and \ref{SQM3MR}). Notice that a {\it stiffer} ($\xi < 0$) SLY can well be consistent with GW170817 and NICER, but it is necessary to constrain this parameter. An interesting study is to consider Bayesian statistics to constrain $\xi$, but this is left to another paper, since it is out of the scope of the present paper.}

\section{Conclusions}\label{sec 4}

{In our previous papers \cite{FA2,FA1}, it is shown that is possible to model spherical stars in $f(T)$ gravity without any perturbative approach or particular assumption to $A$, $B$, $T$, etc., commonly made by some authors in the attempt to simplify the calculations. Also, we have applied this general approach to polytropic EOSs. In the present paper, we extend these studies to a couple of realistic EOSs.} 

{Generally speaking, it is essential to consider realistic EOSs in modeling NSs in any alternative gravity, {since only then it is possible to constrain or ruled out them when confronted with observational data such as gravitational wave data and NICER data. We intend to investigate this restrictions imposed by the experimental data for EOSs as well as the free parameter $\xi$ of the model considered.} {In parallel, Ref. \refcite{lcz} also follows this direction.}}

{Note that different EOS proposals have difficulties in explaining NS with masses above 2 M$_\odot$ in GR, even taking rotation into account. So it is interesting to see what alternative gravity has to say about masses above 2 M$_\odot$.}

As a general conclusion {regarding the $f(T)$ studied here}, one may say that  $\xi > 0$ ($<$ 0) makes the maximum masses smaller (greater) as compared to GR. The greater (smaller) $\xi$ is, the smaller (greater) the maximum mass. These conclusions do not depend on the stiffness of the EOSs. Although, the EOSs studied here are affected slightly differently for a given value of $\xi$.

An interesting conclusion is that depending on the value of $\xi$ it is possible to obtain maximum masses greater than  3 M$_\odot$ even considering soft EOSs. {Recall that soft EOSs are favoured by the GW170817 event (see, e.g., Ref. \refcite{GW170817})}. Therefore, it would be possible to explain the secondary compact object related to the event GW190814.
This is by itself an interesting issue that deserve to be considered in detail. {Additionally, it is important to study the stability of these stars, since a limit mass for $f(T)$ must exist.}

{As we have stressed out along the paper, the calculation of the mass considered here (and also usually by other authors) is exactly the same for GR, although this is under discussion in the literature.}

Last but not least, our next step is to study other $f(T)$ models{, such as $f(T) = T +\xi T^\beta$, where $\xi$ and $\beta$ are free parameters}. In fact, this is already underway and should appear in the literature in due course.

\section*{Acknowledgments}

J.C.N.A. thanks FAPESP $(2013/26258-4)$ and CNPq (308367/2019-7) for partial financial support. The authors would like to thank Rafael da Costa Nunes for discussions related to the $f(T)$ theory. {Last but not least, we thank the referee for the valuable review of our article, which helped to substantially improve our article.}

\end{document}